\documentclass[prb,aps,reprint,amsfonts,amsmath,showpacs]{revtex4-1}

\usepackage{xcolor}
\definecolor{darkgreen}{rgb}{0,0.6,0}
\definecolor{cyan}{rgb}{0,0.7,0.8}
\usepackage[colorlinks=true,citecolor=darkgreen,urlcolor=cyan]{hyperref}
\usepackage{graphicx}
\usepackage{mathrsfs}
\usepackage{bm}

\newcommand{\mrm}[1]{\mathrm{#1}}

\newcommand{\mbb}[1]{\mathbb{#1}}
\newcommand{\mc}[1]{\mathcal{#1}}

\newcommand{\Eref}[1]{Eq.~(\ref{#1})}

\newcommand{\erefr}[2]{(\ref{#1}--\ref{#2})}

\newcommand{\fref}[1]{Fig.~\ref{#1}}

\newcommand{\sref}[1]{Sec.~\ref{#1}}

\newcommand{\pfref}[1]{\protect{Fig.~\ref{#1}}}

\newcommand{\psref}[1]{\protect{Sec.~\ref{#1}}}

\newcommand{\rmi}{\mathrm{i}}
\newcommand{\ra}{\rangle}

\newcommand{\p}[1]{\phantom{#1}}
\newcommand{\rcite}[1]{Ref.~\onlinecite{#1}}
\newcommand{\prcite}[1]{Ref.~\protect{\onlinecite{#1}}}

\newcommand{\rhoA}{\hat\rho_\mrm{A}}
\newcommand{\rhoB}{\hat\rho_\mrm{B}}
\newcommand{\rhoAM}{\hat\rho_A}
\newcommand{\rhoBM}{\hat\rho_B}
\newcommand{\ostar}{\prime}

\begin{document}

\title{Measures of entanglement in non-Abelian anyonic systems}

\author{Robert N. C. Pfeifer}
\email[]{rpfeifer@perimeterinstitute.ca}
\affiliation{Perimeter Institute for Theoretical Physics, 31 Caroline St. N, Waterloo ON~~N2L 2Y5, Canada}

\date{February 5, 2014}

\begin{abstract}
Bipartite entanglement entropies, calculated from the reduced density matrix of a subsystem, provide a description of the resources available within a system for performing quantum information processing. However, these quantities are not uniquely defined on a system of non-Abelian anyons. This paper describes how reduced density matrices and bipartite entanglement entropies (such as the von~Neumann and Renyi entropies) may be constructed for non-Abelian anyonic systems, in ways which reduce to the conventional definitions for systems with only local degrees of freedom.
\end{abstract}

\maketitle

\section{Introduction}

Entanglement represents a fundamental resource for the performance of quantum information processing protocols, corresponding to the existence of non-classical correlations between subsystems. Demonstrations of these non-classical correlations have been performed as long ago as \citeyear{bell1964},\cite{bell1964} and with modern experimental techniques it is becoming increasingly possible to systematically exploit this behaviour.%
\cite{steane1998,bennett2000,ladd2010} 
One area of particular interest is the generation of entanglement in non-Abelian anyonic systems. As well as representing an intriguing and novel phase of matter, these systems are capable of implementing quantum computing protocols with extremely low rates of decoherence, and acting as robust quantum memories.%
\cite{kitaev2003,preskill2004,nayak2008} 
A number of candidates have been proposed for the experimental realisation of such systems.%
\cite{read1999,xia2004,pan2008,kumar2010,stern2010,sanghun-an2011}

An \emph{entanglement monotone} is a mapping $\mc{M}$ between the reduced density matrix description of a physical (sub)system and the real numbers, such that $\mc{M}[L(\hat{\rho})]\leq \mc{M}(\hat\rho)$ for all $\hat\rho$, where $L$ is an arbitrary LOCC (Local Operations and Classical Communication) operation over $\hat\rho$.%
\cite{bennett1996,vidal2001,vedral1997} 
As entanglement cannot be created by LOCC operations but can be eliminated (by collapse of the wavefunction), entanglement monotones provide partial orderings over the space of states, and are sometimes loosely referred to as a \emph{measure of the entanglement} between the system described by $\hat\rho$ and its environment. Multiple such monotones exist.\cite%
{vedral1997,vedral1997a,hamma2005,vidal2000,plenio2007,flammia2009} 

Of particular interest in this paper will be the von~Neumann and Renyi \emph{bipartite entanglement entropies}. On dividing an isolated physical system into two regions A and B, a bipartite entanglement entropy is an entanglement monotone computed from the reduced density matrix of either subsystem A or B (denoted $\rhoA$ and $\rhoB$ respectively) %
which describes the entanglement between regions A and B. 
Specifically, the Renyi entropies comprise a family of entanglement monotones conventionally defined as 
\begin{equation}
S_n(\hat\rho_\mrm{A})=\frac{1}{1-n}\log_2{\left[\mrm{Tr}{\left(\hat\rho_\mrm{A}^{\p{_\mrm{A}}n}\right)}\right]}, \quad n\not=1,\label{eq:Renyi} 
\end{equation}
and the von~Neumann entropy 
\begin{equation}
S_1(\rhoA)=\mrm{Tr}\left(\hat\rho_\mrm{A}\log_2{\hat\rho_\mrm{A}}\right)\label{eq:vonNeumann} 
\end{equation}
corresponds to the limit $n\rightarrow 1$.
A previous attempt has been made to define entanglement monotones for systems of anyons from an information theoretic viewpoint,\cite{hikami2008} with attention primarily to systems of anyons on the disc, and is adequate for Abelian anyonic systems and for appropriately ordered abstract systems of non-Abelian anyons. However, for non-Abelian anyons the formulation becomes ambiguous when applied to real physical systems. The present paper builds on this previous work to %
describe the full family of von~Neumann and Renyi bipartite entanglement entropies for non-Abelian anyonic systems on surfaces of arbitrary genus.

\section{The reduced density matrix\label{sec:rdm}}

In order to compute bipartite entanglement entropies between two regions~A and~B, it is first necessary to construct a reduced density matrix on either region~A or region~B. Although construction of the density matrix for an anyonic system is relatively straightforward, requiring only knowledge of the Hilbert space of the system, the construction of a reduced density matrix for an arbitrary subsystem involves a surprising number of subtleties which impact the subsequent calculation of entanglement entropy. 
Consequently, it is first necessary to discuss in some detail the construction of a reduced density matrix for a subsystem %
admitting
non-Abelian excitations.

\subsection{General properties of the density matrix\label{sec:general}}

Consider first the familiar case of a system of spins $i_1,\ldots,i_n$ on a lattice $\mc{L}$, where the total Hilbert space $\mc{H}$ admits a tensor product structure
\begin{equation}
\mc{H}=\bigotimes_{i=1}^n \mc{H}_i.\label{eq:tproddec}
\end{equation}
The space of operators on the lattice $\mc{L}$ is then given by $\mc{O}=\mc{H}\otimes\overline{\mc{H}}$. The density matrix $\hat\rho$ may be used to compute the expectation value of any operator $\hat{O}\in\mc{O}$ and thus
occupies the space dual to $\mc{O}$ under the action of trace, such that
\begin{equation}
\begin{split}
F_{\hat\rho}&:\mc{O}\longrightarrow\mbb{C}\\
F_{\hat\rho}&(\bullet) = \mrm{Tr}(\hat\rho~\bullet).\label{eq:dmtrace}
\end{split}
\end{equation}
As $\mc{O}$ is self-dual under the action of trace, it follows that $\hat\rho\in\mc{O}$.

For such a system, a subsystem $\mc{A}$ may be defined as a subset of the spins making up $\mc{L}$. If lattice $\mc{L}$ is then placed on a manifold $M$ then there exists a submanifold $A\subseteq M$, possibly disjoint, such that $A$ contains only the spins in $\mc{A}$. The Hilbert space of the portion of the spin system local to submanifold~$A$ is then precisely the Hilbert space of the subset of spins $\mc{A}$,
\begin{equation}
\mc{H}_A = \mc{H}_\mc{A} = \bigotimes_{i\in\mc{A}} \mc{H}_i,
\end{equation}
and it is unnecessary to distinguish between the subset of spins $\mc{A}$ and the enclosing submanifold $A$, both of which may consequently be referred to as ``region~A''. The space of operators on region~A is then $\mc{O}_\mrm{A}=\mc{H}_A\otimes\overline{\mc{H}_A}$. The reduced density matrix $\rhoA$ occupies the space dual to $\mc{O}_\mrm{A}$ under the action of the trace operation,
\begin{equation}
\begin{split}
F_{\rhoA}&:\mc{O}_\mrm{A}\longrightarrow\mbb{C}\\
F_{\rhoA}&(\bullet) = \mrm{Tr}(\rhoA~\bullet)
\end{split}\label{eq:rhoinO}
\end{equation}
and thus also $\rhoA\in\mc{O}_\mrm{A}$.

For a system of non-Abelian anyons, $\mc{O}_\mrm{A}$ will once again be defined as the space of operators local to region~A. Knowledge of the reduced density matrix $\rhoA$ on region~A must permit evaluation of the expectation value of any operator in $\mc{O}_\mrm{A}$, so once again $\rhoA$ must inhabit the space dual to $\mc{O}_\mrm{A}$. However, %
the Hilbert space of a system of non-Abelian anyons does not in general admit the tensor product structure of \Eref{eq:tproddec} and thus the definition of $\mc{O}_\mrm{A}$ must proceed with greater care.

\subsection{The anyonic density matrix\label{sec:adm}}

In preparation for the construction of a reduced density matrix for a subsystem containing non-Abelian anyons, consider first the definition of the density matrix on the entirety of the system.

\begin{figure}
\includegraphics[width=246.0pt]{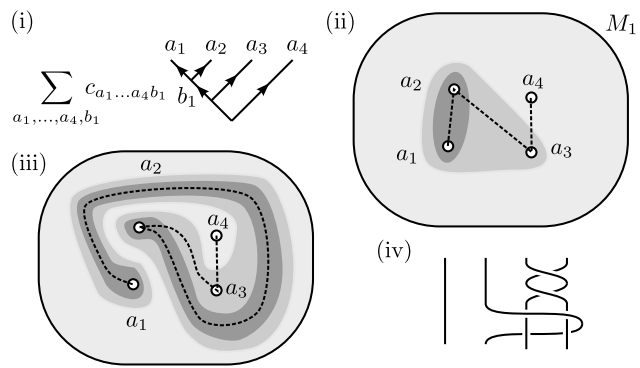}
\caption{When constructing a fusion tree basis on a system of non-Abelian anyons, in order to unambiguously describe the state of a physical system it is necessary to specify the relevant pairs-of-pants decomposition.
For example, when the fusion tree in diagram~(i) is taken in conjunction with the pairs-of-pants decomposition indicated by shading in diagram~(ii), this suffices to specify a state $|\psi_1\ra$ on manifold $M_1$ (which %
is a disc with trivial total charge). If the same sum over weighted fusion trees~(i) is instead associated with the decomposition of $M_1$ given in diagram~(iii), then this specifies a different a different state $|\psi_2\ra$, where $|\psi_2\ra=\hat B_{234}|\psi_1\ra$ for $\hat B_{234}$ given in diagram~(iv). %
In diagrams~(ii) and~(iii), the dashed line represents the projection of the fusion tree onto the manifold when viewed from above, following its initial construction from the pairs-of-pants decomposition but prior to its flattening onto the page in the form of diagram~(i). %
Vertex indices have been suppressed throughout this paper for simplicity, but may easily be reintroduced if required.
\label{fig:projection}} %
\end{figure}

Let a system of anyons inhabit a two-dimensional manifold $M$.
Following the prescription given in \rcite{pfeifer2012a}, %
a fusion tree diagram may be constructed by treating the anyons as punctures and performing a pairs-of-pants decomposition of $M$, then constructing the tree to fit inside the pants. It is well-known that valid labellings of this fusion tree then represent the basis vectors of a Hilbert space describing a system of anyons on $M$ (see e.g. Refs.~\onlinecite{moore1989,pfeifer2012a}); it is less widely appreciated that different pairs-of-pants decompositions may yield fusion trees which appear equivalent but which differ by braiding operations, as illustrated (for example) in \fref{fig:projection}. 
To uniquely specify the state of a physical system it is therefore necessary not only to give a weighted sum over labellings of the fusion tree, but also the relevant pairs-of-pants decomposition. For a surface of genus~0 this may be achieved by specifying the projection of the fusion tree onto the manifold as in \fref{fig:projection}(ii)-(iii), as the pairs-of-pants decomposition may then be reconstructed from the combination of this projection and the fusion tree.

\begin{figure}
\includegraphics[width=246.0pt]{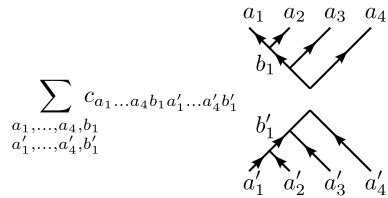}
\caption{Graphical representation of an arbitrary operator $\hat{O}\in\mc{O}_M$ 
for the system of anyons described in \protect{\fref{fig:projection}}(i)-(ii). Projection onto $M_1$ is taken to be as per \pfref{fig:projection}(ii).
\label{fig:operatorM}}
\end{figure}

Having obtained a basis of states $\{|\psi^i_M\ra\}$ which spans the Hilbert space $\mc{H}$ for a system of anyons on an arbitrary manifold $M$, operators acting on $M$ once again inhabit the space $\mc{O}=\mc{H}\otimes\overline{\mc{H}}$, which for anyonic systems will also be denoted $\mc{O}_M$. %
For the example system of \fref{fig:projection}(i)-(ii) this construction is illustrated in \fref{fig:operatorM}. %
The matrix trace of \Eref{eq:dmtrace} is replaced by the (graphical) quantum trace 
\begin{equation*}
\raisebox{-26.5pt}{\includegraphics[width=27.33pt]{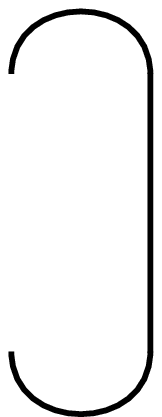}}\textrm{~~~~~~or~~~~~~}\raisebox{-26.5pt}{\includegraphics[width=27.33pt]{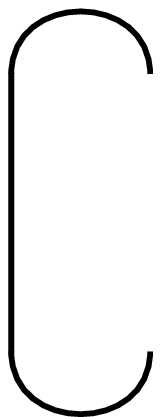}}
\end{equation*}
which is equivalent to the matrix trace on Abelian systems, and the density matrix $\hat\rho$ is once again seen to inhabit $\mc{O}_M$.

\subsection{Non-trivial boundary charges}

The method of constructing a basis for an anyonic system described in \sref{sec:adm} makes use of a powerful duality between Unitary Braided Tensor Categories (UBTCs) and Schwarz-type Topological Quantum Field Theories (TQFTs). If an anyon model is considered as a collection of quasiparticle excitations in a quantum spin liquid, with the nature of the liquid selecting the UBTC which describes the anyon model and the behaviours of these excitations being described by the UBTC using the usual diagrammatic formalism, then the corresponding TQFT description is obtained by identifying the quantum spin liquid with the manifold of the TQFT and replacing the quasiparticle excitations with punctures.

\begin{figure}
\includegraphics[width=246.0pt]{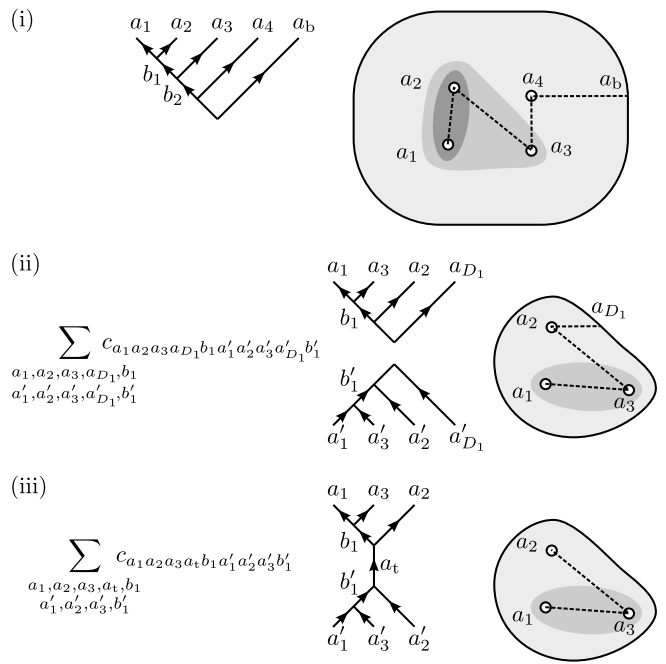}
\caption{(i)~Fusion tree basis for states of four anyons on manifold $M_1$ %
with a possible boundary charge $a_\mrm{b}$.
If $a_\mrm{b}$ is trivial, the corresponding leaf may be deleted from the fusion tree to recover the basis of \fref{fig:projection}(i). Note that the projection of the fusion tree now extends to the boundary of $M_1$.
(ii)~General form of an operator acting on a 3-anyon disc $D_1$ with boundary charge. Label $a_{D_1}$ represents the charge on the boundary of $D_1$. %
(iii)~General form of an operator on $D_1$ which leaves the boundary charge invariant. The total boundary charge is denoted $a_\mrm{t}$, and in this example corresponds to $\overline{a_{D_1}}$ %
for the value of $a_{D_1}$ given in diagram~(ii), 
because the boundary is non-disjoint. Note that the projection of this fusion tree no longer extends to the boundary as the total boundary charge is not associated with a leaf of the fusion tree.
\label{fig:totalcharge}}
\end{figure}

Consequently, whereas the UBTC picture incorporates both anyonic excitations and possible boundary charges of the quantum spin liquid, in the TQFT picture all anyonic charges are associated with a boundary of the manifold. Construction of a fusion tree basis treats all boundaries equivalently, and thus the leaves of the tree may be associated not only with anyons but also with the boundaries of the quantum spin liquid. For example, if the disc of \fref{fig:projection}(ii) were permitted to carry a boundary charge then an appropriate basis would be given by the fusion tree in \fref{fig:totalcharge}(i), as opposed to the one employed in \fref{fig:projection}(i) where the boundary charge was fixed to be trivial.
\begin{figure*}
\includegraphics[width=492.0pt]{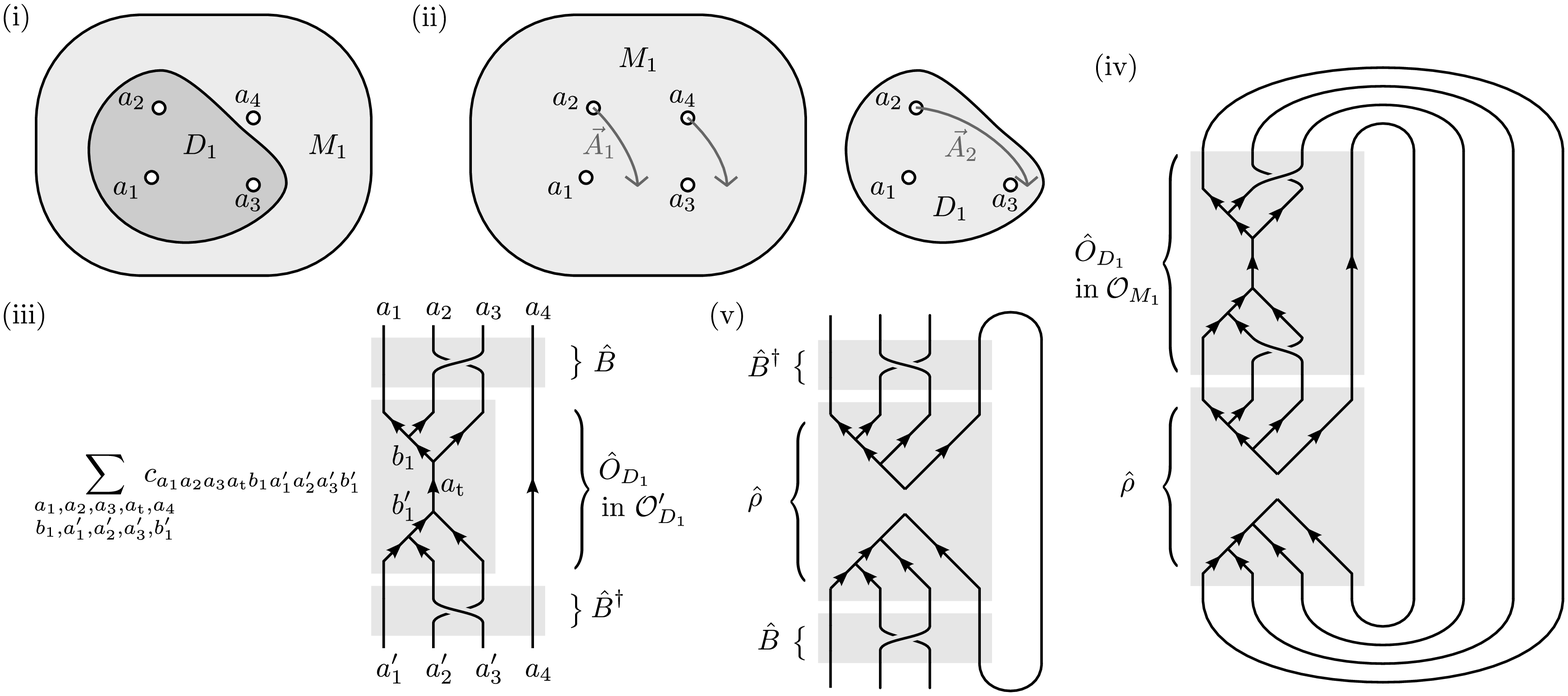}
\caption{(i)~Embedding of $D_1$ in $M_1$. (ii)~The fusion tree bases of \pfref{fig:totalcharge}(iii) and \pfref{fig:operatorM} show the anyons arranged in a straight line. The linearisation of the fusion trees may be achieved on $D_1$ and $M_1$ from the projections given in \pfref{fig:totalcharge}(iii) and \pfref{fig:projection}(ii) respectively by rearranging the relevant anyons as shown. $\vec{A}_1$ and $\vec{A}_2$ label the vectors describing the translations applied to anyon $a_2$ in each case. The net difference in the trajectory of $a_2$ may then be described by vector $\vec{A}_1-\vec{A}_2$, which braids $a_2$ behind $a_3$. (iii)~Consequently, an operator $\hat{O}_{D_1}$ having the form of \pfref{fig:totalcharge}(iii) in $\mc{O}^\prime_{D_1}$ has a representation in $\mc{O}_{M_1}$ as shown [with projection as per \pfref{fig:projection}(ii)]. The coefficients in this expression are the same as in \pfref{fig:totalcharge}(iii), and the mapping between $\mc{O}'_{D_1}$ and $\mc{O}_{M_1}$ is described by the operator $\hat{B}$. (iv)~Given %
a density matrix of the form given in \fref{fig:operatorM}, these diagrams may be concatenated and traced as shown to compute the expectation value of $\hat{O}_{D_1}$. (v)~Construction of a reduced density operator $\hat\rho_{D_1}\!\in\mc{O}^\ostar_{D_1}$ satisfying $\mrm{qTr}(\hat{\rho}_{D_1}\,\hat{O}_{D_1})=\mrm{qTr}(\hat\rho~\hat{O}_{D_1})~\forall~\hat{O}_{D_1}\!\in\mc{O}^\ostar_{D_1}$, where qTr denotes the quantum trace. In diagrams~(iv) and~(v), sums, coefficients, and labels have been suppressed for simplicity.
\label{fig:ODinO}}
\end{figure*}%

As a consequence of this identification between anyons and punctures, it is always possible to identify an anyon model on an $n$-punctured manifold having $m$ disjoint sections of boundary with an anyon model on an equivalent $n+m$-punctured manifold which does not have a boundary. A simple example is the mapping between the $n$-punctured disc and the $n+1$-punctured sphere described in \rcite{pfeifer2012a}.

For any manifold (or, indeed, submanifold) with boundary charges, the Hilbert space of states on that manifold is spanned by the valid labellings of a fusion tree which includes the boundary charges. For example, if disc $D_1$ is a manifold containing three anyons %
$a_1$, $a_2$, and $a_3$, then the space of all possible operators acting on $D_1$ (including those which may change the boundary charge), denoted $\mc{O}_{D_1}$, may be represented by all diagrams of the form of \fref{fig:totalcharge}(ii). %
The space of all operators acting on a general discoid %
manifold $D$ will be denoted $\mc{O}_D$.

It is also useful to describe the space of all operators which do not modify the boundary charge, which will be denoted $\mc{O}_{D}^\ostar$. A diagrammatic basis for $\mc{O}^\ostar_D$ may be obtained by starting with a diagrammatic basis for $\mc{O}_D$ and performing the quantum trace over all charges associated with boundaries of the quantum spin liquid. Regardless of the number of disjoint sections making up the boundary, the result following simplification is a diagram such as that given in \fref{fig:totalcharge}(iii) where a single charge label $a_\mrm{t}$ connects the upper and lower halves of the diagram and represents the total charge on the boundary.

\subsection{The anyonic reduced density matrix\label{sec:ardm}}

\subsubsection{Genus 0\label{sec:genus0}}

Consider next the definition of an operator on a disc $D$ explicitly satisfying $D\subset M$, where $M$ (ignoring any anyon-associated punctures) is a connected manifold of genus~0. %
The space of operators on $D$ is $\mc{O}_D$, and includes operators which modify the boundary charge of $D$. However, on $M$ this boundary is not physical. Consequently, any operator modifying the boundary charge of $D$ must in fact be a truncated description of an operator in $\mc{O}_M$ which transfers charge between $D$ and $M-D$. The action of such an operator on $M-D$ is therefore non-trivial.

Referring back to \sref{sec:general}, it is desireable now to define ``operators local to $D$'' for a system of non-Abelian anyons. Adopting a physical motivation for this definition, an operator is defined as ``local to $D$'' if that operator is trivial everywhere except on $D$. Under this definition, the space of operators local to $D$ corresponds to $\mc{O}_D^\ostar$.

Given specific diagrammatic bases for $\mc{O}_M$ and $\mc{O}^\ostar_D$, comparison of the projection of an operator $\hat{O}_D\in\mc{O}^\ostar_D$ on $D\subset M$ with the projection of an operator $\hat{O}_M\in\mc{O}_M$ on $M$ yields an explicit embedding of $\mc{O}^\ostar_D$ into $\mc{O}_M$. This is illustrated for the example of $D_1\subset M_1$ in \fref{fig:ODinO}(i)-(iii). The embedding of $\mc{O}^\ostar_D$ in $\mc{O}_M$ is defined in terms of a braiding operator $\hat B\in\mc{O}_M$, and knowledge of this operator permits the calculation of expectation values for operators $\hat{O}_D\in\mc{O}^\ostar_D$ using $\hat\rho$ [\fref{fig:ODinO}(iv)].

Applying diagrammatic isotopy to this calculation, %
it is seen that applying the adjoint action $\hat B^\dagger(\bullet)\hat B$ to $\hat\rho$ then tracing out
\begin{enumerate}
\item all anyons not in $D$, and
\item the boundary charges of $D$
\end{enumerate}
yields a reduced density matrix $\hat\rho_D\in\mc{O}^\ostar_D$. %
A basis for $\mc{O}_M$ will be termed \emph{compatible} with a given basis for $\mc{O}^\ostar_D$ if $\hat B$ is trivial.
The construction of $\hat\rho_{D_1}$ for $D_1\subset M_1$ is shown in \fref{fig:ODinO}(v).

\begin{figure*}
\includegraphics[width=492.0pt]{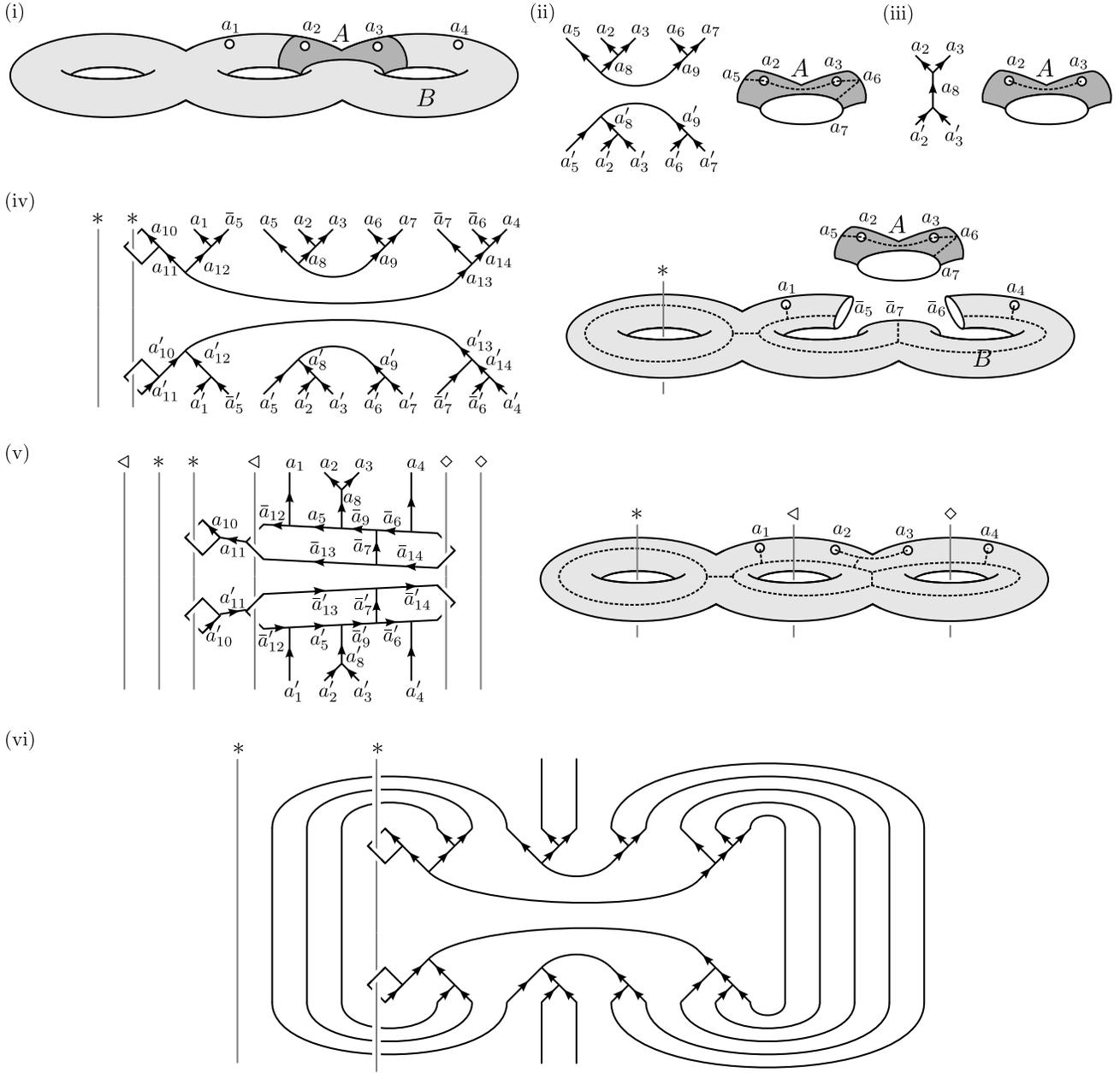}
\caption{Construction of a reduced density matrix for a more complicated system of anyons, using the diagrammatic formalism for surfaces of higher genus given in \prcite{pfeifer2012a}. %
(i)~Manifold $M$ with three handles and four anyons, divided into submanifolds $A$ and $B$. 
(ii)~Choice of fusion tree basis $F_A$ for operators on region $A$, with projection onto $A$.
(iii)~Choice of fusion tree basis $F'_A$ for operators on region $A$, with projection onto $A$.
(iv)~Compatible choice of basis $F_{M'}$ on $M^\ostar$, with projection onto $M'$. On surfaces of nonzero genus, an appropriately chosen projection of the fusion tree may still unambiguously specify the pairs-of-pants decomposition of the manifold when taken in conjunction with the corresponding fusion tree.
(v)~Corresponding fusion tree basis $F_M$ on $M$, with projection.
(vi)~Diagram for calculation of $\hat\rho_A$ from a density operator $\hat{\rho}$ expressed in the basis of diagram~(iv). Sums, labels, and coefficients have been suppressed for clarity. On simplification this reduces to an operator having the form given in diagram~(iii), written in basis $F'_A$ and acting on anyons $a_2$ and $a_3$ but not on the boundary charges of $A$.
In the fusion tree portions of diagrams~(iv)-(vi) the symbols $*$, $\diamond$, and $\triangleleft$ represent identifications in the plane, whereas in the manifold portions of the diagram they indicate the handles which correspond to these identifications, relating the fusion tree to the manifold and fixing the orientation of topological charges such as $a_{10}$. In diagram~(vi) it is unimportant whether traces are performed behind or in front of the identification labelled $*$ so long as this choice is made consistently. Diagrammatic isotopy may be used to move these traces to the right if preferred.
\label{fig:highergenus}}
\end{figure*}%

\subsubsection{Surfaces of higher genus\label{sec:highergenus}}

Generalisation of this approach to %
manifolds $M$ and submanifolds $A$ of arbitrary genus permits the construction of a reduced density matrix $\hat\rho_A$ for any manifold $M$ and submanifold $A\subset M$. This construction may be achieved as follows:
\begin{enumerate}
\item Introduce a fusion tree basis $F_A$ for operators on $A$ which  has leaves for all anyons and all sections of boundary. 
Let $\mc{C}$ be the set of charges associated with boundaries on $A$ which are not also boundaries on $M$ (i.e.~boundaries which would be created on cutting around $A$). Basis $F_A$ must be chosen so that tracing over all charges in $\mc{C}$ may be performed without braiding. The space of operators expressed in basis $F_A$ will be termed $\mc{O}_A$ and is analogous to $\mc{O}_D$ in \sref{sec:genus0}.
\item In basis $F_A$, trace out all boundary charges which are not also associated with a boundary on $M$ (i.e.~all charges in set $\mc{C}$). The resulting space of operators will be termed $\mc{O}_A^\ostar$ and is analogous to $\mc{O}_D^\ostar$ in \sref{sec:genus0}. Let the fusion tree basis of $\mc{O}_A^\ostar$ be denoted $F_A^\ostar$.
\item Let $M^\ostar$ be the disjoint manifold obtained on cutting $M$ around the boundary of $A$. Construct a fusion tree basis $F_{M^\ostar}$ for operators on $M^\ostar$ which is compatible with basis $F_A$.
\item Construct a fusion tree basis $F_M$ for operators on $M$ which yields basis $F_{M^\ostar}$ on $M^\ostar$ when cut around the boundary of $A$.
\item Specify $\hat\rho$ in basis $F_M$.
\item Cut around the boundary of $A$ to compute $\hat\rho$ on $M^\ostar$ in basis $F_{M^\ostar}$.
\item Trace out all anyons not in $A$, all punctures or boundaries created by cutting around the boundary of $A$,\footnote{For situations where the boundary of \protect{$A$} intersects the boundary of \protect{$M$}, a puncture is not traced if any part of its boundary coincides with a boundary on \protect{$M$}. As before, operators coupling the charge on the boundary of \protect{$M$} to other charges in \protect{$A$} will belong to \protect{$\mc{O}^\ostar_A$} if they are trivial on \protect{$M-A$}.} and all degrees of freedom associated with handles in $M$ but not in $A$ (and thus with non-trivial loops in $F_M$ but not in $F_A^\ostar$), to yield $\rhoAM$ in basis $F_A^\ostar$.
\end{enumerate}
An example of constructing a reduced density matrix for a more complicated manifold using this procedure is illustrated %
in \fref{fig:highergenus}.

Note~1: When the genus of submanifold~$A$ is greater than zero, the embedding of $\mc{O}_A^\prime$ into $\mc{O}_{M'}$ in arbitrarily-selected bases $F'_A$ and $F_{M'}$ may involve not only braiding but also transformations acting on topological degrees of freedom. Including these transformations in $\hat{B}$, the definition of compatible bases in Step~3 remains unchanged as $\hat{B}=\mbb{I}$.

Note~2: In the construction given above, the quantum trace is always performed without braiding and without requiring any transformations on the topological degrees of freedom. %
For this to be possible, the density matrix \protect{$\hat\rho$} is first written in a basis $F_M$ which, when cut to yield $\hat\rho$ in basis $F'_M$, is compatible by construction with a basis \protect{$F_A^\ostar$} on \protect{$\mc{O}^\ostar_A$}. %
More generally one may construct a reduced density matrix $\rhoAM$ from a density matrix $\hat\rho$ in \emph{any} basis, including ones not compatible with $F^\ostar_A$, provided the procedure of taking a quantum trace is augmented by a description $\hat{B}$ of the transformations corresponding to the embedding of $\mc{O}'_A$ into $\mc{O}_{M'}$. However, such transformations may always be interpreted as first performing a change of basis on $\hat\rho$ into one %
which \emph{is} compatible with $F^\ostar_A$, followed by performing the quantum trace \emph{without} transforming or braiding. Consequently there is no disadvantage to
requiring that the quantum trace always be performed without braiding or transforming the topological degrees of freedom, having first expressed the density matrix $\hat\rho$ in a basis which, after cutting, will be compatible with $F'_A$.

\subsection{Notes on disjoint manifolds}

It is noted that the computation of $\rhoAM$ presented in \sref{sec:ardm} remains valid even when one or both out of manifold $M$ and submanifold $A$ are disjoint. Like the manifold itself, the fusion tree basis for a disjoint manifold is made up of two or more separate pieces, with one piece corresponding to each disjoint section. %

If a submanifold $A$ is disjoint, then a fusion tree basis on $M$ is said to be compatible with a basis on $A$ if that basis on $M$ is compatible with the basis for each section of $A$ taken independently.

\section{Entanglement entropies\label{sec:entent}}

Given the fusion tree representation of a reduced density matrix, the corresponding matrix representation may be constructed by writing the coefficients in matrix form and absorbing the normalisation factors associated with the fusion tree.%
\cite{bonderson2007,bonderson2008,pfeifer2010}
Once this is done, it is possible to calculate entanglement entropies from the matrix representation of $\rhoAM$ in the usual manner \erefr{eq:Renyi}{eq:vonNeumann}. However, when working with non-Abelian anyons, once again there are additional considerations which must be taken into account.

\subsection{Pairs-of-pants decomposition\label{sec:ententpop}}

It is well-known that braiding operations are capable of generating entanglement between the non-local degrees of freedom of a system of non-Abelian anyons; indeed, this property is exploited when using anyons to implement topological quantum computation.\cite{nayak2008} Furthermore, bases having different projections onto the manifold of the anyonic system, and which therefore linearise the anyons according to different trajectories,
may be related to one another by braiding (see e.g.~Figs.~\ref{fig:projection} and~\ref{fig:ODinO}). 

\begin{figure}
\includegraphics[width=246.0pt]{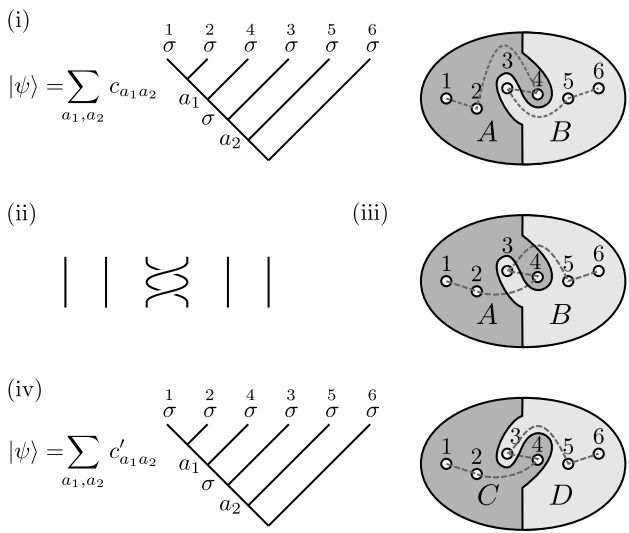}
\caption{(i)~Fusion tree basis for a chain of six Ising anyons on the disc with trivial boundary charge. Note the ordering of anyons on the fusion tree. Fusion products $a_1$ and $a_2$ may either be the fermion charge $\psi$ or the trivial charge $\mbb{I}$. %
(ii)~Braiding operator $\hat{B}_{34}$ affecting entanglement between regions $A$ and $B$. Under the action of this braid, basis elements satisfying $a_1=a_2$ acquire a phase of $\mrm{e}^{-\pi\rmi/4}$ whereas those satisfying $a_1\not=a_2$ acquire a phase of $\mrm{e}^{3\pi\rmi/4}$. This transformation is inherently nonlocal, belonging to neither $\mc{O}'_A$ nor $\mc{O}'_B$, and %
so is capable of generating entanglement between regions $A$ and $B$. 
(iii)~Alternative projection of the fusion tree of diagram~(i) onto the disc, differing from that of diagram~(i) precisely by the braiding operator of diagram~(ii). Note that 
a density matrix in this basis does not admit the construction of a reduced density matrix on region $A$ or region $B$ without first applying an appropriate braiding operator. For example, applying $(\hat{B}_{34})^{-1}$ to diagram~(iii) permits construction of $\rhoAM$ or $\rhoBM$ by recovering the basis of diagram~(i). It is, however, compatible with the construction of a reduced density matrix for region $C$ or region $D$ as shown in diagram~(iv).
\label{fig:sigmabraid}}
\end{figure}

Given a bipartition which divides a manifold $M$ into submanifolds $A$ and $B$ %
and a reduced density matrix on one of these submanifolds, $\hat\rho_X|_{X\in\{A,B\}}$, any unitary operation $\hat{U}_X$ satisfying
\begin{equation}
\hat{U}_X\in\mc{O}'_X 
\end{equation}
will leave the eigenvalues of the reduced density operator $\hat{\rho}_X$ unchanged, as will any operator local to the other submanifold
\begin{equation}
\hat{U}_Y\in\mc{O}'_Y\qquad Y\in\{A,B\},~Y\not=X,
\end{equation}
or any operator which may be reduced to the sequential application of operators of these two types. Such operators will be denoted as belonging to $\mc{O}'_A\times\mc{O}'_B$. In contrast, operators which do not satisfy these criteria
may potentially entangle %
the degrees of freedom associated with regions $A$ and $B$ and thus change the eigenvalues of $\hat\rho_X$. 
A simple example of such an operation is given in \fref{fig:sigmabraid}.

Now let $\mrm{P}_{1A}$ and $\mrm{P}_{1B}$ be pairs-of-pants decompositions of submanifolds $A$ and $B$, and let $\mrm{P}_1$ be a decomposition of $M$ comprising the union of $\mrm{P}_{1A}$ and $\mrm{P}_{1B}$. If $\mrm{P}_2$ is a second such decomposition comprising the union of $\mrm{P}_{2A}$ and $\mrm{P}_{2B}$ for the same submanifolds $A$ and $B$, then because the boundary between submanifolds $A$ and $B$ remains unchanged, the operator $\hat{B}$ mapping between these two bases necessarily admits a decomposition 
\begin{equation}
\hat{B}=\hat{B}_A\times\hat{B}_B,\qquad\hat{B}_A\in\mc{O}'_A,\quad\hat{B}_B\in\mc{O}'_B
\end{equation}
where $\hat{B}_A$ maps between $\mrm{P}_{1A}$ and $\mrm{P}_{2A}$, and $\hat{B}_B$ maps between $\mrm{P}_{1B}$ and $\mrm{P}_{2B}$.%
\footnote{It may be useful to note that the total charge of submanifold $B$ is accessible from within manifold \protect{$A$}. Consequently, operators involving the braid of anyons in \protect{$A$} around the entirety of \protect{$B$}, and vice versa, may be found in \protect{$\mc{O}'_A$} and \protect{$\mc{O}'_B$} respectively.}

In \sref{sec:ardm} it was shown that to compute a reduced density matrix on a general submanifold $A$, it was first necessary to transform the density matrix into a basis permitting the construction of $\rhoAM$. The pairs-of-pants decomposition $\mrm{P}$ associated with such a basis necessarily comprises the union of decompositions $\mrm{P}_{A}$ and $\mrm{P}_{B}$ for submanifolds $A$ and $B$ respectively, and consequently specification of regions $A$ and $B$ suffices to fix the basis up to a unitary transformation in $\mc{O}'_A\times\mc{O}'_B$. A transformation belonging to $\mc{O}'_A\times\mc{O}'_B$ may not modify the eigenvalues of the reduced density matrix, and therefore when computing a bipartite entanglement entropy it is unnecessary to give a specific pairs-of-pants decomposition of the manifold.

It does remain necessary, however, to specify the decomposition of manifold $M$ into submanifolds $A$ and $B$. This is clearly seen in \fref{fig:sigmabraid} where diagrams~(i) and~(iv) differ by the entanglement-generating operation of \fref{fig:sigmabraid}(ii) but admit fusion tree bases distinguishable only in terms of their projection onto the manifold. Merely stating that the bipartition separates the anyons into two groups \{1,2,4\} and \{3,5,6\} fails to distinguish between these two distinct scenarios with different entanglement entropies.

\subsection{Summary of considerations}

Systems of non-Abelian anyons admit larger families of entanglement monotones than do systems governed by Abelian statistics.
When computing a bipartite entanglement entropy for non-Abelian anyons inhabiting a two-dimensional manifold $M$, in order to fully specify which entanglement entropy has been computed it is necessary to explicitly specify the bipartitioning of the manifold, not just the separation of the anyonic excitations into two groups.
It is not, however, necessary to provide a specific pairs-of-pants decomposition of the manifold.

In contrast, the pairs-of-pants decomposition \emph{must} be given in order to unambiguously describe a state or operator on a physical system, including the density matrix. This consideration has, in the past, been frequently overlooked for anyon chains as the ordering of the anyons on the fusion tree and the natural ordering of anyons on the manifold (e.g.~\{1,2,3,4,5,6\} in \fref{fig:sigmabraid}) are assumed to coincide. The mapping between the fusion tree and the pairs-of-pants decomposition of the manifold attains greater significance in systems which are not quasi-one-dimensional, and in contexts which force a choice of basis that does not correspond to the natural ordering of the chain. An example of this is the construction of a reduced density matrix suitable for computing bipartite entanglement entropies between regions~$A$ and $B$ in \fref{fig:sigmabraid}(i).

\section{Discussion}

The study of entanglement entropies in topologically ordered systems %
is currently a hot topic%
\cite{castelnovo2007,chen2010,eisert2010,levin2006,furukawa2007,hamma2008,haque2007,jiang2012,zhang2011,nussinov2009,papanikolaou2007,li2008,kato2013} 
with a major focus being the calculation of the topological entanglement entropy\cite{kitaev2006a} as an aid to the classification of different topologically ordered phases of matter. Fortunately this quantity depends only upon the topology of the manifold and the nature of the anyon model describing the quasiparticle excitations,
and thus even in studies of non-Abelian anyons (e.g.~\rcite{kato2013}) its calculation is unaffected by the considerations raised in the present discussion. As interest increases in the condensed matter physics of anyonic systems, however, attention will turn to the behaviours of the entanglement monotones themselves, and it is vital that these quantities be well-defined if successful study is to be made of entanglement in systems of non-Abelian anyons. %

Previous work on the subject\cite{hikami2008} has concentrated on defining entanglement entropy with respect to bipartitions on an abstract fusion tree, separating the anyons into two groups (denoted $A$ and $B$) without attention to the manifold, and requiring that the
anyons from both groups be contiguous on that tree. For the study of real physical systems, this definition is insufficient as the relationship of the physical system to the fusion tree is ambiguous, resulting in bipartite entanglement entropies which are not well-defined. To resolve this ambiguity it is necessary to specify the bipartition of the system in terms of the manifold rather than simply grouping the anyons on the fusion tree.
\begin{figure}
\includegraphics[width=246.0pt]{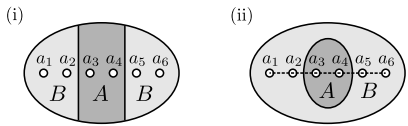}
\caption{The prescription given in \psref{sec:ardm} permits calculation of entanglement entropies between the centre and edge of a chain of anyons even when anyons in one region are not contiguous on the fusion tree. This is essential when working with disjoint submanifolds, as in diagram~(i), and may also simplify the calculation of entanglement entropies for central regions when employing a naturally-ordered fusion tree basis on the chain, as shown in diagram~(ii).
\label{fig:expandingregion}}
\end{figure}%

Furthermore, it has proved possible to relax the requirement that anyons from both regions~$A$ and~$B$ be contiguous on the fusion tree. The most stringent requirements of contiguousness arise when a contiguous manifold $M$ is divided into two contiguous submanifolds $A$ and $B$. Even then, in order to permit construction of bipartite entanglement entropies it suffices
that the anyons inhabiting at least one of these two regions (but not necessarily both) be consecutive in the fusion tree on $M$%
. %
If constructing the reduced density matrix on submanifold $A$, then relaxing the requirement that anyons in submanifold $B$ be consecutive in the fusion tree basis of $M$ leads to a construction for bipartite %
entanglement entropies not considered in \rcite{hikami2008}. This construction is essential for computing entanglement entropies where one submanifold is made up of two disjoint parts as shown in \fref{fig:expandingregion}(i), and may also be convenient when comparing entanglement between the centre and ends of a chain %
as shown in \fref{fig:expandingregion}(ii).

It is also interesting to compare the current work with that of \citeauthor{kato2013},\cite{kato2013} in which entanglement entropies for non-Abelian anyonic systems are computed by embedding these systems into the Hilbert space of a spin system (see also \rcite{pfeifer2012a}). In choosing a basis in which to perform this embedding, and in maintaining a notion of locality during the mapping from the manifold to the spin system, the embedding process and bipartition on the spin system may be understood to implicitly define submanifolds $A$ and $B$ on the original manifold (up to the equivalence class of deformations which are trivial when expressed in the fusion tree basis). %
Consequently the work of these authors is entirely consistent with, and may be understood in terms of, the framework presented herein.

Finally, it is noted that the considerations raised in the present paper are unique to non-Abelian anyons. When applied to systems of fermions, bosons, or Abelian anyons, the definitions presented above always reduce to the usual definitions of entanglement entropies for these systems.

The author thanks the Ontario Ministry of Research and Innovation Early Researcher Awards for financial support. Research at Perimeter Institute is supported by the Government of Canada through Industry Canada and by the Province of Ontario through the Ministry of Research and Innovation.

\bibliography{EntEnt}

\end{document}